\documentclass[fonts]{icst}

\usepackage{moreverb}

\usepackage[breaklinks,colorlinks,bookmarksopen,bookmarksnumbered,linkcolor=ICSTblue,citecolor=blue,urlcolor=ICSTblue]{hyperref}
\usepackage{breakurl}

\usepackage{amssymb}
\usepackage{graphicx}
\usepackage[ruled,vlined,linesnumbered]{algorithm2e}
\usepackage{float}
\usepackage{color}
\usepackage{soul}
\usepackage{amsmath}
\usepackage{caption}
\usepackage{makecell}

\usepackage{url}

\newcommand\BibTeX{{\rmfamily B\kern-.05em \textsc{i\kern-.025em b}\kern-.08em
T\kern-.1667em\lower.7ex\hbox{E}\kern-.125emX}}

\def\volumeyear{2014}

\journalname{XXXXXX}
\articletype{Research Article}
 \def\volumeyear{XXXX}
 \def\volumenumber{XX}
  \def\issuemonth{XX-XX}
  \def\issuenumber{X}
  \def\articleid{XX}
  \copyrightnote{This is an open access article distributed under the terms of the Creative Commons Attribution license (\url{http://creativecommons.org/licenses/by/3.0/}), which permits unlimited use, distribution and reproduction in any medium so long as the original work is properly cited.}
  \def\articledoi{10.4108/XX.X.X.XX}
\received{XXXX}
  \accepted{XXXX}
  \published{XXXX}

\begin{document}

\runningheads{Iqbal H. Sarker}{Mobile Data Science: Towards Understanding Data-Driven Intelligent Mobile Applications}

\title{Mobile Data Science: Towards Understanding Data-Driven Intelligent Mobile Applications}

\author{Iqbal H. Sarker \affil{*}}

\address{Department of Computer Science and Software Engineering, \\ Swinburne University of Technology, \\ Melbourne, VIC-3122, Australia.}

\abstract{Due to the popularity of smart mobile phones and context-aware technology, various contextual data relevant to users' diverse activities with mobile phones is available around us. This enables the study on mobile phone data and context-awareness in computing, for the purpose of building data-driven intelligent mobile applications, not only on a single device but also in a distributed environment for the benefit of end users. Based on the availability of mobile phone data, and the usefulness of data-driven applications, in this paper, we discuss about \textit{mobile data science} that involves in collecting the mobile phone data from various sources and building data-driven models using machine learning techniques, in order to make dynamic decisions intelligently in various day-to-day situations of the users. For this, we first discuss the \textit{fundamental concepts} and the \textit{potentiality} of mobile data science to build intelligent applications. We also highlight the key \textit{elements} and explain various key \textit{modules} involving in the process of mobile data science. This article is the first in the field to draw a big picture, and thinking about mobile data science, and it's potentiality in developing various data-driven intelligent mobile applications. We believe this study will help both the researchers and application developers for building smart data-driven mobile applications, to assist the end mobile phone users in their daily activities.}

\keywords{Mobile phone user, smartphone data, data science, behavioral analytics, mobile data mining, machine learning, data-driven decision making, contexts, context-awareness, ambient intelligence, intelligent mobile services, mobile systems and applications, pervasive computing, intelligent environment.}

\tnotetext[1]{This article is a preprint version of the journal \textit{"EAI Endorsed Transactions on Scalable Information Systems"}. \\}

\fnotetext[1]{Corresponding author: Iqbal H. Sarker.  Email: \email{msarker@swin.edu.au}}

\maketitle

\section{Introduction}
\begin{figure*}[htbp!]
	\centering
	\includegraphics[width=.8\linewidth]{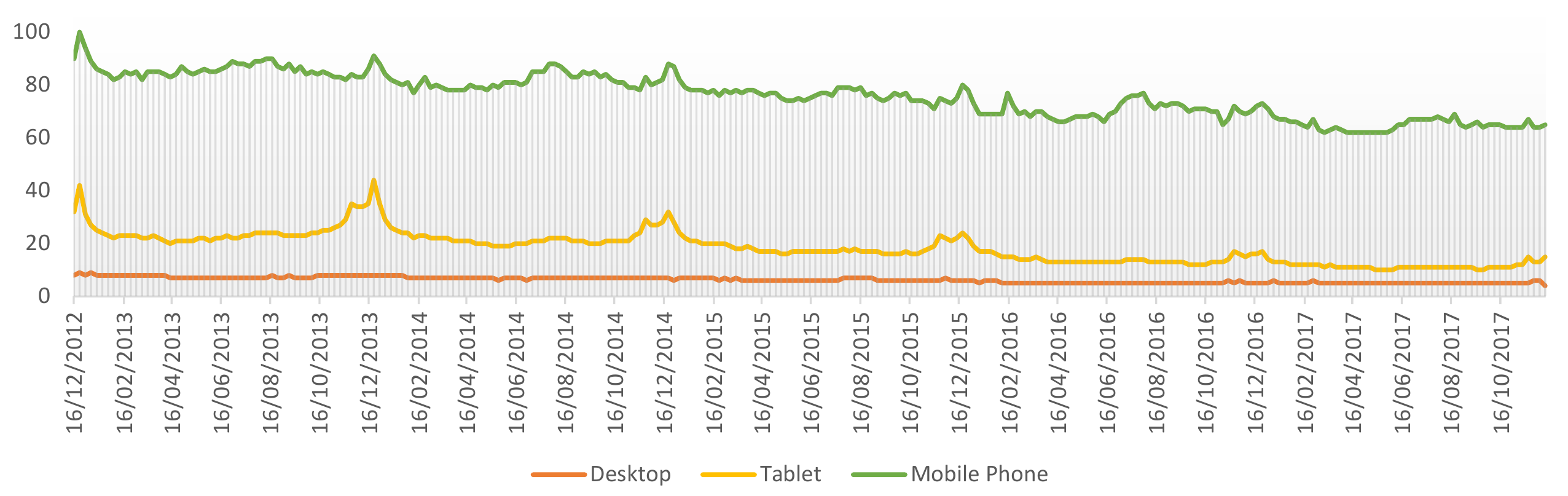}
	\caption{Users' interest trends over time}
	\label{fig:interest-trends}
\end{figure*}

The advancement of mobile computing and the Internet have played a significant role in the development of the current digital age. The Internet has now formed the backbone of modern communication. Nowadays, the use of the Internet, particularly the World Wide Web (WWW) has moved beyond desktop computers to millions of mobile phones. According to ITU (International Telecommunication Union), cellular network coverage has reached 96.8\% of the world population, and this number even reaches 100\% of the population in the developed countries \cite{number-of-mobile-phone-users}. With the development of science and technology, the smart phone industry is growing rapidly and the popularity of smart phones has made exponential growth in mobile phone application market \cite{peng2018personalized}. According to Google Trends \cite{GoogleTrends2017}, users' interest on \textit{``Mobile Phones''} is more and more than other platforms like \textit{``Desktop Computer''} or \textit{``Tablet Computer''} over time. For instance, Figure \ref{fig:interest-trends} shows users' interests trends over time on mobile phones over desktop and tablet computer for the last five years (December 2012 to December 2017). The data was collected on 13th December 2017 from Google Trends \cite{GoogleTrends2017}. 

In Figure \ref{fig:interest-trends}, the numbers from 0 to 100 in Y-axis, represent search interests in terms of popularity relative to the highest point on the chart for a given date (e.g., 16/12/2012). For instance, a value of 100 represents the peak popularity for a particular term. A value of 50 means that the term is half as popular. Likewise, a score of 0 means the term was less than 1\% as popular as the peak \cite{GoogleTrends2017}. Based on the popularity of mobile phones over time shown in Figure \ref{fig:interest-trends}, we are motivated to think about \textit{mobile data science} and corresponding data-driven intelligent mobile applications to assist the end mobile phone users in their daily activities.

Nowadays, mobile phones have become one of the primary ways, in which people around the globe communicate with each other. These devices have transformed over a period of time from merely communication tools to smart and highly personal devices. Such devices are able to assist us in a variety of day-to-day situations in our daily activities. People in the world use mobile phones for various activities, such as for voice communication, Internet browsing, using mobile applications (apps), e-mailing, using online social network, instant messaging etc \cite{pejovic2014interruptme}. While mobile phones may come in various forms, in this paper they refer to smart mobile phones or Internet accessible mobile phones which have incorporated many advanced features to facilitate better information access and utilization for the benefit of end mobile phone users in their daily life. These smartphones in addition to being used as a communication device are capable of doing things that a computer does because of their computing capability on the devices. For instance, what actually makes the mobile phone ``smart''? The answer is it's ability to handle data, not only the voice calls. In general, these devices can fulfill nearly all communication requirements of the users in recent times. Smart mobile phones are not only used as communication devices but also used in various sectors \cite{lane2010survey} like health care services, navigation services, business purposes, safety, environmental monitoring, intelligent transport systems, traffic management, intelligent route routing, smart homes, or smart cities etc.

In recent times, the smartphones are becoming more and more powerful in both computing and data storage aspects. Their more and more sensing capabilities have enabled the collection of rich raw contextual information of the user, surrounding real-life environment, and the mobile phones usage records through the device logs, such as phone call logs \cite{phithakkitnukoon2011behavior} \cite{sarker2016phone}, SMS Logs \cite{eagle2006reality}, mobile application (Apps) usages logs \cite{zhu2014mining} \cite{srinivasan2014mobileminer}, mobile phone notification logs \cite{mehrotra2016prefminer}, web logs \cite{halvey2005time}, game Logs \cite{paireekreng2009time}, context logs \cite{cao2010effective} \cite{zhu2014mining}, and smartphone life logs \cite{rawassizadeh2013ubiqlog} etc. The logged data generated by the smart phones provides a means to get new knowledge about various aspects of the users, like user social interactions, various user activities with the devices, which offers the potential to understand the insight of such mobile phone data.

Typically, the traditional data science helps to solve problems by analyzing data and turns data into data products \cite{cao2017data} \cite{han2011data}. However, the concept of mobile data science is not exactly the same thing. The purpose of mobile data science is slightly different, as the smart mobile phones are aware of their user's real-life surrounding environment, and users' various types of activities or social interactions, in various contexts in the real world. Thus, \textit{mobile data science} is a term which is concerned with the challenge of collecting user's real life data in different contexts, such as temporal, spatial, social or others, in a pervasive computing environment, finding data-driven models in order to make dynamic decisions for individual mobile phone users based on relevant contextual information, and use this discovered knowledge for building smart mobile applications to intelligently assist themselves in their daily life. Some examples of such real-life applications are context-aware smart mobile communication, intelligent mobile notification management systems, context-aware smart searching, context-aware mobile recommender system, context-aware app management system, and various context-aware predictive services etc. Thus, the concept of \textit{``mobile data science''} involving users' real-life activities with mobile phones, corresponding contextual information of the users, and the surrounding dynamic environment, has become interested.

In particular, the contributions of this paper are: 

\begin{itemize}
	\item We explore the potentiality of mobile data science for building various data-driven intelligent mobile applications.
	
	\item We review various kinds of mobile phone data and define mobile data science that includes mobile phone data having contextual information and the processing to discover useful knowledge for data-driven applications.
	
	\item We discuss the fundamental concepts of mobile data science and highlight the key differences between the traditional data science and mobile data science in various aspects.
\end{itemize}

The rest of the paper is organized as follows. We briefly review the context-aware mobile computing, which is related to mobile data science, in Section \ref{Contexts and Context-Aware Mobile Computing: A Background}. In Section \ref{Mobile Data Science}, we briefly discuss about mobile data science including the fundamental concepts, and the key modules of mobile data science. Finally, Section \ref{Conclusion} concludes this paper.

\section{Context-Aware Mobile Computing: A Background}
\label{Contexts and Context-Aware Mobile Computing: A Background}
The term \textit{context} can be used with a variety of different meanings in different purposes. In this section, first we briefly discuss about various definitions of contexts in the area of mobile and pervasive computing, and then we discuss about \textit{context-awareness} in mobile computing.

\subsection{Definitions of Contexts}
The notion of context has been used in numerous areas, including Pervasive and Ubiquitous Computing, Human Computer Interaction, Computer-Supported Collaborative Work, and Ambient Intelligence \cite{dourish2004we}. In the area of Ubiquitous and Pervasive Computing, a number of early works on context-aware computing or context-awareness referred context as the locational contexts, i.e., the location of people and objects \cite{schilit1994disseminating}. In recent works, context has been extended to include a broader collection of factors in addition to such locational context, such as physical and social aspects of an entity or object, as well as the activities of users \cite{dourish2004we}. Having examined the definitions and categories of context given by the pervasive and ubiquitous computing community, in this section, we review the definition of contexts within the scope of this paper.

Several studies have attempted to define and represent context from different
perspectives. For instance, the user's location information, the surrounding people and objects around the user, and the changes to those objects are considered as contexts by Schilit et al. \cite{schilit1994disseminating}. Brown et al. \cite{brown1997context} also define contexts as user's locational information, temporal information, the surrounding people around the user, temperature, etc. Similarly, the user's locational information, environmental information, temporal information, user's identity, are also taken into account as contexts By Ryan et al. \cite{ryan1999enhanced}.

Other definitions of context have simply provided synonyms for context such as context as the environment or social situation. A number of researchers are taken into account the context as the environmental information of the user. For instance, in \cite{brown1995stick}, the environmental information that the user's computer knows about are taken into account as context by Brown et al, whereas the social situation of the user is considered as a context in Franklin et al. \cite{franklin1998all}. On the other hand, a number of other researchers consider it to be the environment related to the applications. For instance, Ward et al. \cite{ward1997new} consider the state of the surrounding information of the applications as contexts. Hull et al. \cite{hull1997towards} define context as the aspects of the current situation of the user and include the entire environment. The settings of applications are also treated as context in Rodden et al. \cite{rodden1998exploiting}. In \cite{schilit1994context}, Schilit et al. claim that the important aspects of context are: (i) where you are, (ii) whom you are with, and (iii) what resources are nearby. The information of the changing environment is taken into account as context in their definition. In addition to the user environment (e.g., user location, nearby people around the user, and the current social situation of the user), they also include the computing environment and the physical environment. For instance, connectivity, available processors, user input and display, network capacity, and costs of computing can be the examples of the computing environment, while the noise level, temperature, the lighting level, can be the examples of the physical environment.

Dey et al. \cite{dey2001understanding} presents a survey of alternative view of context, which are largely imprecise and indirect, typically defining context by synonym or example. Finally, Dey et al. \cite{dey2001understanding} offer a definition of context, which is perhaps now the most widely accepted. According to Dey et al. \cite{dey2001understanding} \textit{``Context is any information that can be used to characterize the situation of an entity. An entity is person, place or object that is considered relevant to the interaction between a user and an application, including the user and the application themselves''}. 

\begin{table*}[htbp!]
	\begin{center}
		\caption{Various Types of Contexts}
		
		\label{tab:context-examples}
		\begin{tabular}{l|c} 
			\textbf{Context Category} & \textbf{Context Examples}\\
			\hline
			User Identity & \makecell{Name, ID, gender, age, date of birth, nationality, \\ educational level, available income etc.}\\
			\hline
			
			Temporal & \makecell{Date (e.g., YYYY-MM-DD), time (e.g., hh:mm:ss), \\ periods (e.g., 1 hour), weekday (e.g., Monday), \\ weekend (e.g., Saturday), season (e.g., summer), \\ public holidays (e.g., Christmas day), etc.}\\
			\hline
			
			Spatial & \makecell{Location (e.g., office, home, work, market, etc.), \\ orientation, speed, etc.}\\
			\hline
			
			Environmental & \makecell{Temperature, weather, humidity, light, \\ sound, noise, traffic conditions, etc.}\\
			\hline
			
			Social & \makecell{Interpersonal relationship (e.g., friend, family, professionals etc.), \\situation (e.g., meeting, lecture, seminar, etc.), \\people nearby, activity change, calendar, etc.}\\
			\hline
			
			Physiological & \makecell{Blood pressure, heart rate, \\ tone of voice etc.}\\
			\hline
			
			Psychology & \makecell{Emotions (e.g., happy, sad, angry etc.) \\ preference, tiredness etc.}\\
			\hline
			
			Mobile phone activity & \makecell{Phone calls, using mobile applications, \\ mobile web and navigation, gaming, using social network \\ (e.g., Facebook, Linkedin) etc.}\\
			\hline
			
			Physical activity & \makecell{Sitting, standing, driving, walking, biking, \\ running, etc.}\\
			\hline
			
			Device-related & \makecell{Illumination, carrier, Wi-Fi,
				Bluetooth, \\ screen status, battery status, setting status, \\ and
				device status etc.}\\
			\hline
			
			App-related & \makecell{Running apps, active app, app status etc.}\\
			\hline
		\end{tabular}
	\end{center}
\end{table*}

\subsection{Context-Awareness}
Typically, a desktop computer application expects a static execution environment, either in office, home or other locations. However, this precondition is generally not applicable for mobile services or systems, as the world around an application is changing frequently. Thus, the nature of mobile applications must change according to the movement of the users. In particular, mobile applications should become more flexible in order to respond in highly dynamic computing environments, which makes the computing more pervasive. Recent advances in smart mobile phones and the pervasive computing environment make it possible in the real-world, where \textit{context-awareness} can be used as the spirit of pervasive computing \cite{shi2006context}. The primary objective of pervasive computing environment is the creation of an environment saturated with seamlessly integrated devices with computing and communication capabilities having decision-making ability with as little direct user interaction as possible \cite{anagnostopoulos2005context}. The use of contextual information
in mobile applications is able to reduce the amount of human effort and attention that is needed for an application to provide the services according to user's needs or preferences, in a pervasive computing environment. Thus, context-awareness simply represents the dynamic nature of the applications.

Nowadays, the popularity of context-awareness in mobile computing is increasing because of their adaptation in dynamic environment. Typically, context-awareness is originated as a term from ubiquitous or pervasive computing, which is able to deal with linking changes in the real-world environment with mobile systems. According to Wikipedia \cite{wikiContextAwareness}, context-awareness enables a new class of applications in pervasive computing, which is a property of mobile devices that is defined complementarily to location awareness, i.e., adapting capability in the applications with the movement of mobile phone users. However, as discussed above, various types of contexts can be applied to make the mobile applications more flexible and useful according to users' overall situations in terms of temporal, spatial or social, and their preferences, as context-awareness is the main spirit of pervasive computing to build adaptive applications. In general, context-aware computing refers to a general class of mobile systems that can sense their surrounding physical environment, and able to adapt their behavior intelligently according to the sensed information in the corresponding applications \cite{wikiContextAware-computing}.

Context-aware systems are a component of a ubiquitous computing or pervasive computing environment \cite{wikiContextAware-computing}. According to \cite{wikiContextAware-computing}, there are three important aspects of context; these are: where you are; who you are with; and what resources are nearby. In order to make the mobile applications capable of operating in highly dynamic environments demanding on less user attention, different types of contexts might have different impact, shown in Sarker et al. \cite{sarker2017anapproach}. Hence, we summarize a number of contexts, related to mobile phone users, their activities, and surrounding environment, shown in Table \ref{tab:context-examples}.

\section{Understanding Mobile Data Science}
\label{Mobile Data Science}
We are living in the age of data science \cite{cao2017data}. On the other hand, computing is increasing being carried out using smart mobile phones, which support not only the telephony but also data-driven mobile applications. In this section, we discuss briefly about the mobile data science, ranging from smartphone raw data having contextual information, building data-driven models using machine learning techniques, and corresponding smart mobile applications for the end users.

\subsection{Motivational Examples}
The following real-life examples intuitively illustrate the advantage of \textit{mobile data science} in various mobile applications either on single device or in a distributed environment, to assist the users in their daily life. In general, a single device based application is related to the computation on individuals' devices utilizing their phone log data. On the other hand, application in the distributed environment is designed to allow users of a computer network to access information, and services, as well as to exchange information with others, like a client-server based model.

EXAMPLE 1 (Application Scenario on Single Device). Say, Alice, a smartphone user, is a PhD student. She has installed a large number of mobile applications (apps) on her smartphone. Homescreens of smartphones provide ready access to commonly used apps, which is particularly useful. However, the homescreen of her smartphone is unaware about Alice's changing contexts, such as her location, and consequently unable to manage apps intelligently according to her needs. Thus, an intelligent mobile app management system is needed for her, which allows the particular app she currently needs to be easily accessible from the mobile homescreen. To build such intelligent applications \textit{mobile data science} can play an important role by predicting her future usages utilizing her app usages log data, that intelligently assists her to use different types of mobile apps, such as Skype, Whatsapp, Facebook, Gmail, Youtube, Linkedin, Microsoft Outlook, etc. according to her current contextual information.

EXAMPLE 2 (Application Scenario in Distributed Environment). Let's consider a cloud-based mobile service recommendation system for the above user Alice. When she enters her University campus by driving her car, she will automatically get a recommendation for the `best' car parking space to minimize the searching time and cost. In addition, the user will receive on her mobile device detailed driving instructions to reach that space. A distributed cloud-based system is able to accomplish this job automatically without any additional effort of the user. She can get this service using the corresponding client application installed on her mobile phone based on her contextual information and preferences. The cloud-based system is typically a client-server model where in the mobile device based agent (client) extracts the real-time contextual information of the user and communicates with a cloud-based service that houses the corresponding recommender system (server). Unlike the single device based services, where computation is done within the device, the battery life of the device, cpu computational limitations, and user preference modeling limitations, can be ignored in a cloud based services as the server is responsible for all the computations needed. To build such cloud based services \textit{mobile data science} can play an important role to predict the parking availability according to her current contextual information by analyzing the relevant data in the cloud.

The above scenarios show that \textit{mobile data science} is potential to build various data-driven intelligent mobile applications, to intelligently assist the users in various purposes by analyzing the relevant data, not only on a single device but also in a distributed environment.

\subsection{Understanding Real-Life Smartphone Data}
Real-life smartphone data having contextual information relevant to individual mobile phone users' activity is one of the key elements of mobile data science. The reason is that mobile phone data with contextual information is the basis of mobile data science to build data-driven intelligent applications for the users. According to \cite{cao2017data}, we live in the age of data, where everything that surrounds us is linked to a data source and everything in our lives is captured digitally. On the other hand, mobile or cellular phones have become increasingly ubiquitous and powerful. Recent advances in smart mobile phones and their sensing capabilities have enabled the collection of users various activities and corresponding contextual information. Thus, contextual data is available around us to analyze. In this section, we discuss about a variety of mobile phone data containing the associated contextual information of individual users. We have summarized these data in Table \ref{tab:mobile-phone-data}.

\begin{table*}[htbp!]
	\begin{center}
		\caption{Various Types of Smart Mobile Phone Data}
		\label{tab:mobile-phone-data}
		\begin{tabular}{l|c|c} 
			\textbf{Smartphone Data} & \textbf{\makecell{Data Description \\(user activity and contexts information)}} & \textbf{References}\\
			\hline
			Call Log & \makecell{User phone call activities, \\ such as incoming, missed and outgoing calls, \\ and corresponding contextual information \\ such as date, time, call type, \\ call duration, user identifier, location, etc.} & \makecell{Sarker et al. \cite{sarker2017individualized}, \\ Eagle et al. \cite{eagle2006reality}, \\ Srinivasan et al. \cite{srinivasan2014mobileminer}, \\ Bell et al. \cite{bell2011nodobo}}\\
			\hline
			
			SMS Log & \makecell{Spam and non-spam text messages \\ and related contextual information such as \\ user identifier, date, time, etc.} & \makecell{Almeida et al. \cite{almeida2011contributions}, \\ Eagle et al. \cite{eagle2006reality}}\\
			\hline
			
			App Usages Log & \makecell{User various app usages such as \\ Gmail, Skype, Facebook, Linkedin, Read news, \\ Whatsapp, SNS, Web, Multimedia, Game, etc.\\ and corresponding contextual information such as \\ date, time-of-the-day, battery level, profile type \\ (General, Silent, Meeting, Outdoor, Pager, Offline), \\ Charging state (Charging, Complete, Not Connected), \\ location (Home, Work Place, On the way) etc.} & \makecell{Zhu et al. \cite{zhu2014mining}, \\ Srinivasan et al. \cite{srinivasan2014mobileminer}}\\
			\hline
			
			Notification Log & \makecell{Users' response for various notifications, \\ such as promotional emails, game invites on social network, \\  predictive suggestions by applications, etc. \\ and corresponding contextual information such as \\ date, time-of-the-day, notification type, user's physical activity \\ (includes still, walking, running, biking and in vehicle), \\ location (Home, Work, or Others) etc.} & \makecell{Mehrotra et al. \cite{mehrotra2016prefminer}}\\
			\hline
			
			Web Log & \makecell{User mobile web navigation, web searching, \\ e-mailing, entertainment, chat, music, news, \\ TV, netting, traveling, sport, banking, etc. \\ and related contextual information such as \\ date, time-of-the-day, weekdays, weekends, etc.} & \makecell{Halvey et al. \cite{halvey2005time},\\ Halvey et al. \cite{halvey2006time}}\\
			\hline
			
			Game Log & \makecell{Users' various game playing such as action, \\ adventure, casual, puzzle, RPG, strategy, sports, etc. \\ and related contextual information such as \\ date, time-of-the-day, weekdays, weekends, etc.} & \makecell{Paireekreng et al. \cite{paireekreng2009time}}\\
			\hline
			
			Smartphone Life Log & \makecell{User phone calls, SMS headers (no content), \\ App usage (e.g., Skype, Whatsapp, Youtube etc.), \\ physical activities form Google play API, \\ and related contextual information such as  \\ WiFi and Bluetooth devices in user's proximity, \\ geographical location, temporal information, etc.} & \makecell{Rawassizadeh et al. \cite{rawassizadeh2013ubiqlog}}\\
			\hline
			
		\end{tabular}
	\end{center}
\end{table*}

\begin{table*}[h!]
	\begin{center}
		\caption{Some Key Terms in Mobile Data Science}
		
		\label{tab:key-terms}
		\begin{tabular}{c|c} 
			\textbf{Key Terms} & \textbf{Description}\\
			\hline
			\makecell{Context} & \makecell{Any information that can be used to characterize \\ the situation of an entity (e.g., temporal, spatial or social contexts.)}\\
			\hline
			
			\makecell{Pervasive computing \\ environment} & \makecell{An intelligent environment which is dynamic.}\\
			\hline
			
			Context-awareness & \makecell{The spirit of pervasive computing, \\ which represents the dynamic nature of mobile applications.}\\
			\hline
			
			\makecell{Context-aware \\ mobile computing} & \makecell{A mobile computing paradigm in which applications can discover and \\ take advantage of contextual information.}\\
			\hline
			
			Data science & \makecell{The science of data; also known as data-driven science, \\ is an interdisciplinary field about scientific methods, processes, and systems \\ to extract knowledge or insights from data, similar to data mining.}\\
			\hline
			
			Mobile user activity & \makecell{Physical activity (e.g., walking, driving, etc.) and \\ mobile phone activity (e.g., phone calling, apps using, \\ emailing, Internet browsing, messaging, etc.)}\\
			\hline
			
			Adaptability & \makecell{Able to adapt dynamic environment and behave accordingly \\ (e.g., context-awareness in mobile applications.)}\\
			\hline
		\end{tabular}
	\end{center}
\end{table*}

\begin{table*}[htbp!]
	\begin{center}
		\caption{Traditional Data Science VS Mobile Data Science}
		
		\label{tab:mobile-data-science}
		\begin{tabular}{l|c|c} 
			\textbf{Aspects} & \textbf{Data Science} & \textbf{Mobile Data Science}\\
			\hline
			Rationale & \makecell{Data tells the story} & \makecell{Mobile phone data with contextual information \\ and pervasive intelligence disclose \\ problem-solving solutions for building mobile applications.}\\
			\hline
			Objective & \makecell{Innovative and effective \\ algorithms, Creating  \\ data products} & \makecell{Solving real-life problems of mobile phone users, \\ To build adaptive, intelligent and \\ context-aware mobile software systems, \\ To assist mobile phone users \\ in their daily activities.}\\
			\hline
			Data & \makecell{General purpose data} & \makecell{Mobile phone users's real-life data \\ and surrounding environment information \\ collected by mobile phones \\ in a pervasive computing environment, \\ Data includes contextual information \\ and mobile phone users' various activities.} \\
			\hline
			Mechanism & Automated & \makecell{Human-centered, \\ particularly mobile phone users.}\\
			\hline
			Evaluation & Technical metrics & \makecell{Trade-off between technical significance \\ and mobile phone users' expectation according to \\ their current contexts and preferences.}\\
			\hline
		\end{tabular}
	\end{center}
\end{table*}

These context-rich historical mobile phone data shown in Table \ref{tab:mobile-phone-data}, is simply as the collection of the past contexts and user's actions for the past contexts \cite{hong2009context}. The main characteristic of such kind of log data is that they contain the actual behavior of the users in different contexts, as the mobile phones automatically record these data. Thus, it is important to study mobile phone data involving contextual information and users diverse activities in different contexts, for the purpose of building data-driven smart and context-aware intelligent mobile applications.

\subsection{Fundamental Concepts of Mobile Data Science}
The traditional data science \cite{cao2017data} can be treated as a part of mobile data science. Nowadays, many researchers use of the term ``data science'' to describe the interdisciplinary field of data collection, preprocessing, inferring or making decisions by analyzing the data. According to Cao et al. \cite{cao2017data}, a high-level statement about data science is: ``it is the science of data or the study of data''. The outputs of data science are typically data products depending on what type of data are taken into account, which can be a discovery, prediction, service, recommendation, thinking, model, or system. According to the definition of Cao et al. \cite{cao2017data}, the ultimate values of data products are knowledge, intelligence, wisdom, and decision based on relevant data.

On the other hand, computing is moving toward pervasive, ubiquitous environments \cite{finin2001information}. Thus, the computing devices, software agents, and services are all expected to seamlessly integrate and cooperate in support of human objectives according to their needs and preferences. Mobile devices are able to provide such service in a pervasive computing environment in an anywhere, any-time fashion. Therefore, the next step for pervasive computing is the integration of smart devices that are able to understand the local context of the users and share this information in support of human needs intelligently.

Based on the potentiality of data science and the pervasiveness in computing using mobile devices discussed above, in this paper, we introduce the concept of \textit{mobile data science}, by taking advantages from both, in order to help the researchers and application developers, for the purpose of building data-driven mobile applications utilizing relevant mobile phone data. Our objective is to highlight the importance of working in this area, in order to identify examples of key research issues for the community, thereby acting as a catalyst for new research and development for the benefit of end mobile phone users. We define the field as follows:

\textit{``Mobile data science is research or working area that exists at the intersection of context-aware mobile computing and data science, characterized by a focus on the real-life contextual data collection, data pre-processing, analyzing data using data mining and machine learning techniques, and use of extracted knowledge from the mobile phone data, for the purpose of building smart data-driven mobile applications, in order to intelligently assist the end mobile phone users by providing dynamic decisions in their daily activities, in a context-aware pervasive computing environment.''}.

\begin{figure}[htbp!]
	\centering
	\includegraphics[width = .9\linewidth]{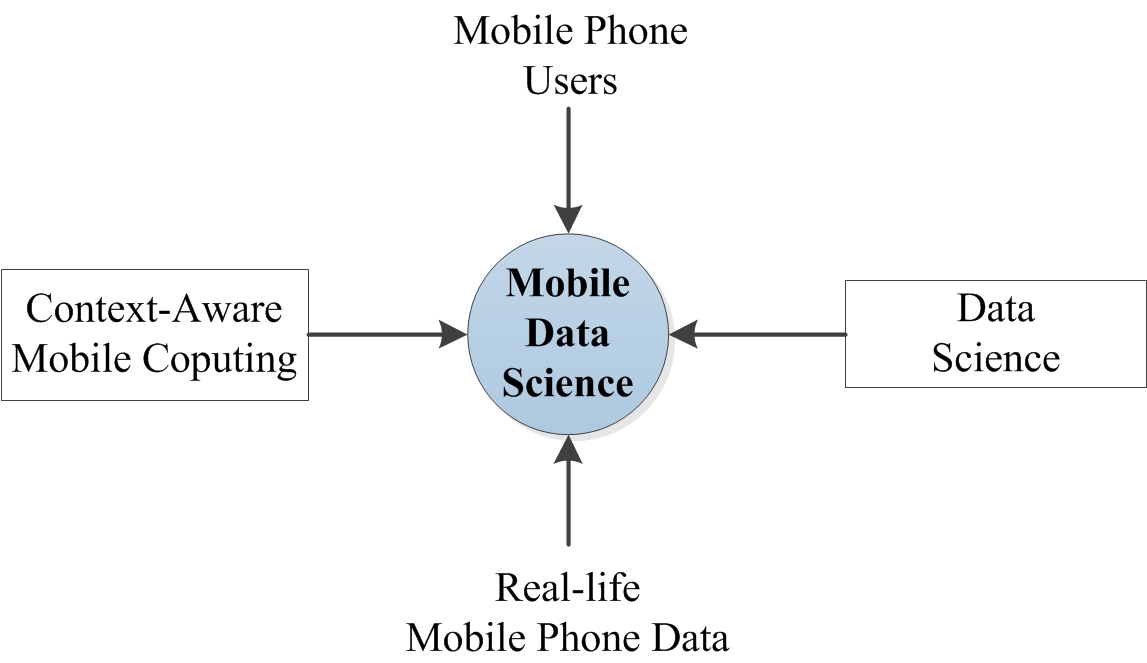}
	\caption{Key involvement in mobile data science}
	\label{fig:overview}
\end{figure}

In Table \ref{tab:key-terms}, we define some key terms related to \textit{mobile data science}. In a word, mobile data science is related to users' real-life mobile phone data, mobile user activities, context-aware mobile computing, and the traditional data science, shown in Figure \ref{fig:overview}. We also summarize various aspects of \textit{mobile data science} over the traditional data science, highlighted in Table \ref{tab:mobile-data-science}.

\subsection{Key Modules of Mobile Data Science}
In this section, we briefly discuss about the key \textit{modules} that are involved in the process of mobile data science, for the purpose of building data-driven intelligent mobile applications. 

\begin{figure}[htbp!]
	\centering
	\includegraphics[width =.9\linewidth]{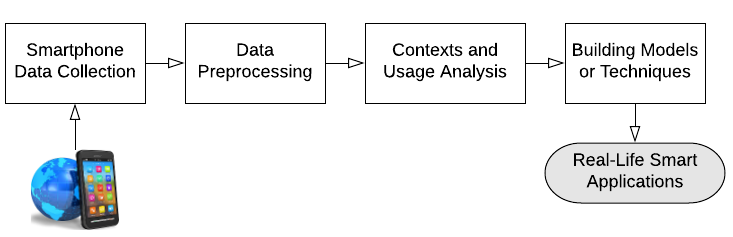}
	\caption{A high-level overview of mobile data science process involving the key modules.}
	\label{fig:overview-process}
\end{figure}

In the process of mobile data science, first the contextual data is collected from the real world using the devices. After that, a data-driven model based on machine learning techniques according to the target problem, can be built by analyzing and processing the contextual data. Such data-driven models can be used to build various intelligent mobile applications to intelligently assist the end mobile phone users in their daily activities. Based on this process, Figure \ref{fig:overview-process} shows a high level overview of mobile data science involving a number of modules, from real-world raw data to applications. In the following, we discuss these modules one by one with examples.

\begin{itemize}
	\item \textit{Mobile Data Collection}: This is the first module of mobile data science, as mobile phone data is the basis of mobile data science. Thus, this module mainly focuses on collecting user's real life smartphone data relevant to a particular problem for analysis and making corresponding decisions. Real-world mobile phone data usually comprise features whose interpretation depends on some \textit{contextual information} such as temporal (e.g., in the morning), spatial (e.g., at office), social contexts (e.g., in a meeting) or others. These contextual-sensitive features and their available patterns in the dataset are of high interest to be discovered and analyzed in order to make dynamic decisions intelligently in a pervasive computing environment. The contextual data can be collected from various sources like smartphone sensors, e.g., GPS, Wi-Fi, Blue-tooth, accelerometer, proximity, camera, microphone, light sensors, etc. or using corresponding phone log data, electronic calendars, according to the needs for a particular application. For instance, Sarker et al. \cite{sarker2016behavior} collect mobile phone call log data including time-series contextual information for their temporal analysis in order to build time-dependent call interruption management system based on users' unavailability \cite{sarker2018Unavailability}.
	
	\item \textit{Data Pre-processing}: This module mainly focuses on preparing the data for modeling. For instance, cleaning the data to address the data quality issues can be a pre-processing task. Real-world mobile phone data may contain noisy or inconsistence instances, missing values, duplicate records, invalid data or outliers. Such noisy instances may have impact on machine learning techniques to build an effective model for a particular problem. Therefore, handling these inconsistency to ensure the quality of data is an important issue for data analysis. Various machine learning techniques can be used to detect such outliers or noisy instances. For instance, in \cite{sarker2017improved}, the authors propose a naive Bayes classifier based noise detection technique to ensure the quality of smart phone data. In addition to the quality of data, to transform the raw data to the formatted or desired data to make it suitable for further analysis is another task in data pre-processing. For instance, the continuous time-series mobile phone data is converted into behavior-oriented segments to capture the behavioral patterns of individual mobile phone users in \cite{sarker2017individualized}. Like the temporal context, the data-centric social context \cite{sarker2018DataCentricSocialContext} or others relevant raw data may need to transform into the desired data that can be used for further analysis according to the needs.
	
	\item \textit{Contexts and Usages Analysis}: This module mainly focuses on selecting the interesting features or contexts that have an influence on users' diverse activities to build an effective model. For instance, social contexts, e.g., in a meeting, or social relationship between individuals, e.g., boss, might have an influence to make a call handling decision of an individual mobile phone users \cite{sarker2018BehavMiner}. Similarly, other contexts such as temporal, or spatial contexts, can play a role in different applications depending on users' need and preferences \cite{sarker2017designing}. For instance, building a context-aware model for predicting parking availability in smart cities, temporal and spatial contexts can be used \cite{zheng2015parking}. Such relevancy of contexts may differ from application-to-application in the real world.
		
	\item \textit{Building Data-driven Models}: Once the required contextual information and mobile phone usages are prepared according to a particular problem, this module is responsible to build an intelligent data-driven model to solve the target problem. Thus, this module can be treated as the core module in mobile data science, as intelligent decision making is depended on the corresponding data-driven model. For the purpose of building such models, various popular data mining and machine learning techniques \cite{han2011data}, such as classification analysis (e.g., Decision tree, Naive Bayes, Support vector machine, K-nearest neighbors, Artificial neural network etc.), ensemble learning (e.g., Random forest, AdaBoost, Bootstrap aggregating etc.), clustering (e.g., K-means, Agglomerative, Divisive etc.), regression analysis (e.g., Linear regression, Logistic regression etc.), and association analysis (e.g., AIS, Apriori, FP-growth etc.) or others relevant machine learning techniques \cite{han2011data} \cite{witten2005data}, can be used for an effective modeling according to the specific needs for a particular problem. The extracted knowledge using such machine leaning techniques then can be used to build various real-life smart mobile applications to intelligently assist the end users. For instance, mining behavioral association rules of individual mobile phone users based on machine learning technique, can be used to effectively predict their future behavior according to their current contextual information \cite{sarker2018mining}.
\end{itemize}

The modules discussed above are the key in mobile data science for the purpose of building data-driven intelligent mobile applications, in order to assist the end mobile phone users in their daily activities in a context-aware pervasive computing environment.

\section{Conclusion}
\label{Conclusion}
In this paper, we have briefly discussed about the concept of mobile data science by highlighting the characteristics of mobile phone data including users' various activities in different contexts. In particular, we focused our investigation on context-awareness and real-life mobile phone data, and their proper utilization for building intelligent data-driven models not only on the single devices but also in a distributed environment, to assist them in their daily activities in a pervasive computing environment. Additionally, we have summarized the key differences between the traditional data science and mobile data science in various aspects. We believe this article will help both the researchers and application developers for building various data-driven intelligent mobile applications for the benefit of end mobile phone users utilizing their mobile phone data.

\section*{Acknowledgment}
\label{Acknowledgment}
I would like to thank Prof. Jun Han, Swinburne University of Technology, Australia, Dr. Alan Colman, Swinburne University of Technology, Australia, Dr. Ashad Kabir, Charles Sturt University, Australia, for their relevant discussions.

\bibliographystyle{plain}
\bibliography{bibfile/PhDThesisBib,bibfile/mobile-data-science,bibfile/MyPapers}
\end{document}